\documentclass[runningheads]{llncs}
\usepackage{graphicx}
\usepackage[thinlines]{easytable}
\usepackage{multirow}
\begin{document}
\title{Attacks on Visualization-Based Malware Detection: Balancing Effectiveness and Executability}
\author{Hadjer Benkraouda\inst{1}\orcidID{0000-0001-5511-3182} \and
Jingyu Qian\inst{1}\orcidID{0000-0002-3953-5382} \and
Hung Quoc Tran\inst{1}\orcidID{0000-0001-7767-3180}\thanks{Benkraouda, Qian, and Tran share co-first authorship.}\and
Berkay Kaplan\inst{1}\orcidID{0000-0002-4365-7606}}
\authorrunning{Hadjer et al.}
\institute{University of Illinois at Urbana-Champaign, Urbana, IL 61801-2302, USA}
\maketitle              % typeset the header of the contribution
\begin{abstract}
With the rapid development of machine learning for image classification, researchers have found new applications of visualization techniques in malware detection. By converting binary code into images, researchers have shown satisfactory results in applying machine learning to extract features that are difficult to discover manually. Such visualization-based malware detection methods can capture malware patterns from many different malware families and improve malware detection speed. On the other hand, recent research has also shown adversarial attacks against such visualization-based malware detection. Attackers can generate adversarial examples by perturbing the malware binary in non-reachable regions, such as padding at the end of the binary. Alternatively, attackers can perturb the malware image embedding and then verify the executability of the malware post-transformation. One major limitation of the first attack scenario is that a simple pre-processing step can remove the perturbations before classification. For the second attack scenario, it is hard to maintain the original malware's executability and functionality. In this work, we provide literature review on existing malware visualization techniques and attacks against them. We summarize the limitation of the previous work, and design a new adversarial example attack against visualization-based malware detection that can evade pre-processing filtering and maintain the original malware functionality. We test our attack on a public malware dataset and achieve a 98\% success rate. 

\end{abstract}
\section{Introduction}

With the proliferation of connectivity and smart devices in all aspects of human life, these devices have become increasingly targeted by malicious actors. One of the most common forms of attack is through malicious software (i.e., malware). According to AVTEST, one of the leading independent research institute for IT security, to date, there have been more than 100 million new malware applications in 2020 alone~\cite{avtest}. Inspired by the success of machine learning in other fields, researchers have proposed using machine learning for many security applications. With the rapid growth and evolving nature of new malware applications, machine learning-based solutions are a natural fit for malware detection and classification due to their robustness. 

Several papers have designed malware detection systems using machine learning (e.g.,~\cite{defense1,defense2}). The proposed solutions tackle malware detection from different perspectives. The main differences are in the representation used and the subsequent machine learning model selected for effective classification. These representations include raw bytes, embeddings, representative features, and binary visualization. Visualization methods, in particular, have shown high accuracy in detecting malware compared to conventional methods. These data reduction and visualization techniques for detection have shown improvements in both speed and memory efficiency ~\cite{speed,efficiency}. Additionally, visualization-based techniques have achieved higher detection accuracy, mainly attributed to the applicability of deep learning techniques in detecting malware patterns~\cite{DeepL}. We, therefore, focus our work on visualization-based malware detection models.

Nevertheless, machine learning models are susceptible to adversarial example attacks, which add imperceptible non-random perturbations to test samples, causing machine learning models to misclassify them. Successful adversarial examples have been seen to fool systems into misclassifying people~\cite{face-rec}, cause systems to recognize street stop signs as speed limit signs~\cite{stop}, or cause voice-controllable systems to misinterpret commands or perform arbitrary commands~\cite{voice}. 

Recent work has shown that machine learning-based techniques for malware detection are also susceptible to adversarial examples. In these systems, the attacks alter the malware binaries to cause the target model to classify the malware sample as benign or vise versa. However, adversarial examples in this domain are more challenging to produce. In addition to the constraint of imperceptibility and minimal changes that conventional adversarial examples must comply with, adversarial examples for malware detection must maintain the original malware functionality. This means that the attacker cannot change the bytes arbitrarily. Instead, the attacker has to understand the functionality of the malware and perform careful changes.

There have been previous attempts to create adversarial examples against visualization-based malware classifiers~\cite{khormali2019copycat,liu2019atmpa}. These attacks produce adversarial examples either by using conventional image-based techniques such as the Fast Gradient Sign Method~\cite{FGSM} or Carlini and Wagner method ~\cite{carlini2017towards} or by injecting byte values to unreachable regions within the binary. These attacks are simplistic and can be detected and removed easily with minimal countermeasures~\cite{removal}.

\begin{figure}[h]
\centering
\includegraphics[width=\linewidth]{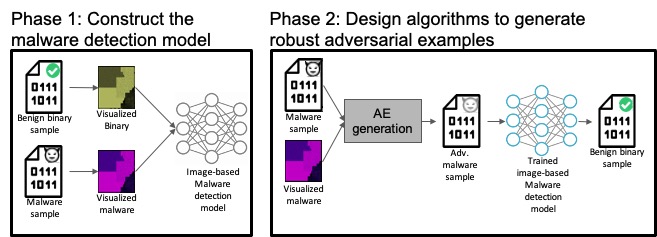}
\caption{The two project phases: constructing the malware detection model and designing the adversarial example attack.}
\label{f:overview}
\end{figure}

In this paper, we propose a new adversarial example attack that combines binary rewriting and adversarial attacks in image classification. We target a convolutional neural network (i.e., CNN) model for malware detection. Because there is no open-sourced code for visualization-based malware detection, our first phase of the project includes constructing the malware detection model (Figure \ref{f:overview} left). We apply a similar CNN structure as previous work for visualization-based malware detection and achieves an overall accuracy of 99\%. In the second phase of the project (Figure \ref{f:overview} right), we design an adversarial example attack against this malware detection model. Our attack performs additive changes to the original malware and ensures that the added instructions are semantic NOPs, i.e., they do not change values of any register or manipulate the program state. Our attack achieves a 98\% success rate on a public malware dataset. The success of the proposed attack reveals that it is necessary for visualization-based malware detection to perform more advanced and robust protection against adversarial examples other than simply filtering the padding or the non-reachable header section. 

\noindent The rest of the paper is organized as follows. In Section \ref{sec:background}, we introduce background and related work on visualization-based malware detection and adversarial machine learning. In Section \ref{sec:attack}, we provide our detailed design of the attack against visualization-based malware detection and illustrate how we solve the challenges of creating successful adversarial examples while maintaining the original malware functionality. In Section \ref{sec:eval}, we discuss our experiment setup and measure our attack success rate. In Section \ref{sec:discussion}, we discuss limitations of our attack and potential future work. 

\section{Background and Related Work}
\label{sec:background}

In this section, we introduce the background and related work on visualization-based malware detection. We then discuss some traditional methods to camouflage malware, adversarial machine learning and how attacks against image classification and malware detection work. Finally, we include a table for SoK of the papers that we mentioned, and point out their limitations. 

\subsection{Malware Visualization}
With the development of image processing technology, visualization-based techniques are also proposed for malware detection and analysis. These techniques can be applied directly to the binary without complicated disassembly and execution process. Researchers have proposed approaches to visualize malware as gray-scale or RGB-colored images. From these images, machine learning techniques, such as CNN, can classify whether the tested software is benign or malicious. 

\begin{figure}[h]
\centering
\includegraphics[width=\linewidth]{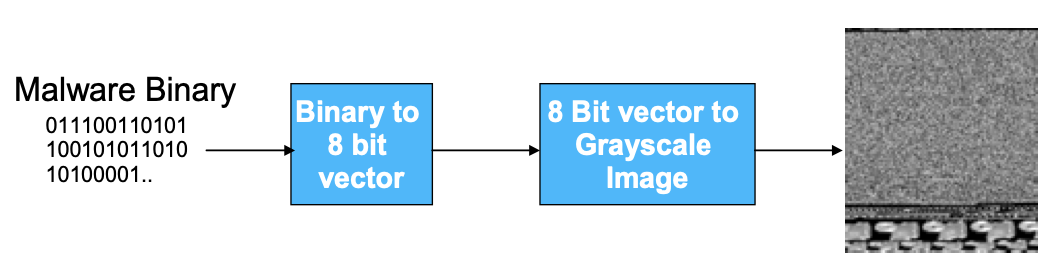}
\caption{Typical approach to visualize the binary in gray-scale~\cite{nataraj2011malware}}
\label{f:bin_to_image}
\end{figure}

Figure \ref{f:bin_to_image} illustrates a typical approach to visualize the malware as an image. The malware binary is grouped by 8-bit vectors. Each vector represents a value from 0 to 255, which can be mapped to a gray-scale pixel value. The shape of the final image depends on the width of the image, which is usually a tunable parameter, and the size of the malware binary in bytes. This methodology can be adapted to visualize the malware as an RGB-colored image, which considers different feature types and represents them in different color channels~\cite{fu2018malware}.

\noindent Here we introduce a few projects from the literature that used several different malware visualization techniques. Han et al. \cite{han2015malware} proposed a malware visualization method that converts the binary to a gray-scale bitmap image and then generates the entropy graph for the entire malware binary. They then used a histogram similarity measuring method to group malware within the same malware family. Nataraj et al. \cite{nataraj2011malware} also visualized the malware binary as a gray-scale image but extracted texture features to characterize and analyze the malware. They used GIST to capture texture features, and apply the K-nearest neighbors algorithm with Euclidean distance to classify the malware into different malware families. Xiaofang et al. \cite{xiaofang2014malware} mapped malware binaries to gray-scale images, extracted a 64-dimension feature vector from the image, and performed fingerprint matching to identify similar malware.

\begin{figure}[h]
\centering
\includegraphics[width=\linewidth]{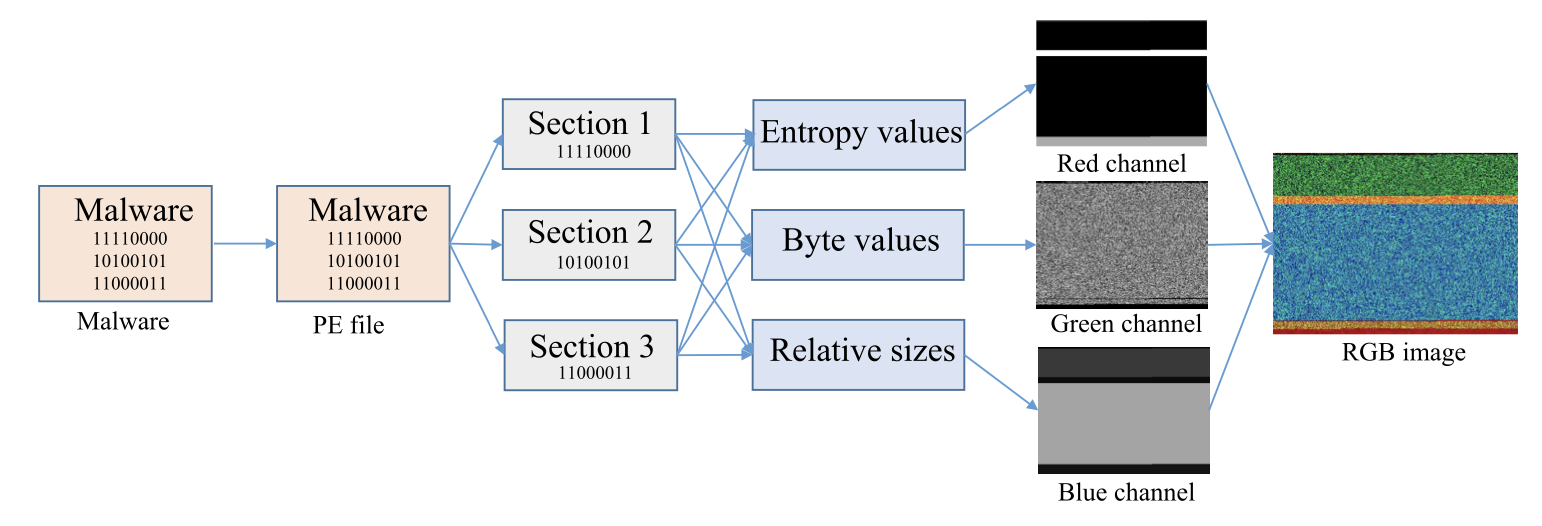}
\caption{Visualize the malware as an RGB image, considering both local and global features~\cite{fu2018malware}}
\label{f:bin_to_rgb}
\end{figure}

Unlike the other work, which converts the malware binary to a gray-scale image, Fu et al. \cite{fu2018malware} took a different approach to visualize the malware (Figure \ref{f:bin_to_rgb}). Their approach extracted both local and global features to generate an RGB-colored image for the malware. Specifically, they extracted three types of features, including section entropy, raw byte values, and relative size of each section to the whole file. For raw byte values, they use the same approach to visualize malware as gray-scale images (Figure \ref{f:bin_to_image}). Each types of features occupies a single color channel (i.e., either red, green, or blue). For the final classification process, they compared different machine learning techniques, including random forest, K-nearest neighbors, and support vector machine.

Han et al.~\cite{han2013malware} proposed a novel method to classify malware and malware families. They extracted opcode instruction sequences from the malware binary as features and generated the RGB-colored image matrix from these features. The image matrices are compared with each other using selective area matching~\cite{han2013malware} to group malware into malware families. 

Another work interestingly extends the field by visualizing the behavior of the malware instead of the malware binary itself, and suggests that any feature of a malware can be visualized to determine its classification~\cite{shaid2014malware}. This work further indicates that the possibility of malware visualization is limitless as a program has a lot of useful features ranging from its behavior to its metadata. But, for this work, it specifically focuses on the malware behavior by running an API call monitoring utility through a Virtual Machine (i.e., VM) to examine the APIs used while the program is executed in user mode~\cite{shaid2014malware}. While there are several other techniques to capture malware behavior, such as capturing the network activity of the malware, or the changes in the operating system's resources, API monitoring has been chosen in this study due to its conciseness and the shortcomings of other techniques that have been discussed in detail~\cite{shaid2014malware}. Afterwards, the calls are mapped to hot colors, specifically such as red, or orange, for classification. Finally, the classification is mostly done through a similarity ratio to the tested software against the known malware's color mapping~\cite{shaid2014malware}.

%Besides other techniques related to the visualization of certain malware files' features, further work in malware analysis using its binary file includes a project that maps image matrices generated from the binary file to RGB values \cite{han2013malware}. Although this type of mapping and the use of binary file as the main feature seems to be common in the field, we assume that its high success rate has led to an abundance of projects in the literature, which also motivated us to use the binary file as the main feature of the malware due to its simplicity. To the best of our knowledge, most projects also focused on similarity ratios to classify programs as malicious or benign. The mentioned study also used a similarity ratio for its classification task after the mapping of the image matrix is created~\cite{han2013malware}. However, we thought that the ratio itself would be insufficient, and more advanced techniques would be required to perform the classification task.

\subsection{Traditional Malware Camouflage}

Although there are several methods of malware detection based on visualization, attackers can still employ various methods such as encryption and obfuscation to hide their malware in the targeted software's code and counter static malware detection methods~\cite{christodorescu2005semantics,liu2017automatic,rad2012camouflage}. 

Malware encryption intends to encrypt the malware body to hide its intentions and avoid static analysis detection so that a direct signature matching defense cannot detect the malware~\cite{christodorescu2005semantics,liu2017automatic,islam2013classification}. It relies on a decryption loop (a.k.a., decryptor) to decrypt the malicious payload and execute the malware. Therefore, if the defense can find out the decryption loops in the malware, it can decrypt the malicious code first and then perform a simple signature matching to detect the malware. In addition, the original defense can be easily augmented with a signature checking component to identify suspicious decryption loops within the malware. Visualization-based defense can be even better at detecting malware encryption because it can extract locality features specific to suspicious decryption loops of the malware. Even a more complicated malware encryption technique that picks different decryptors for different malware (i.e., oligomorphism) only prolongs the detection time~\cite{rad2012camouflage}.

\begin{figure}[h]
\centering
\includegraphics[width=\linewidth]{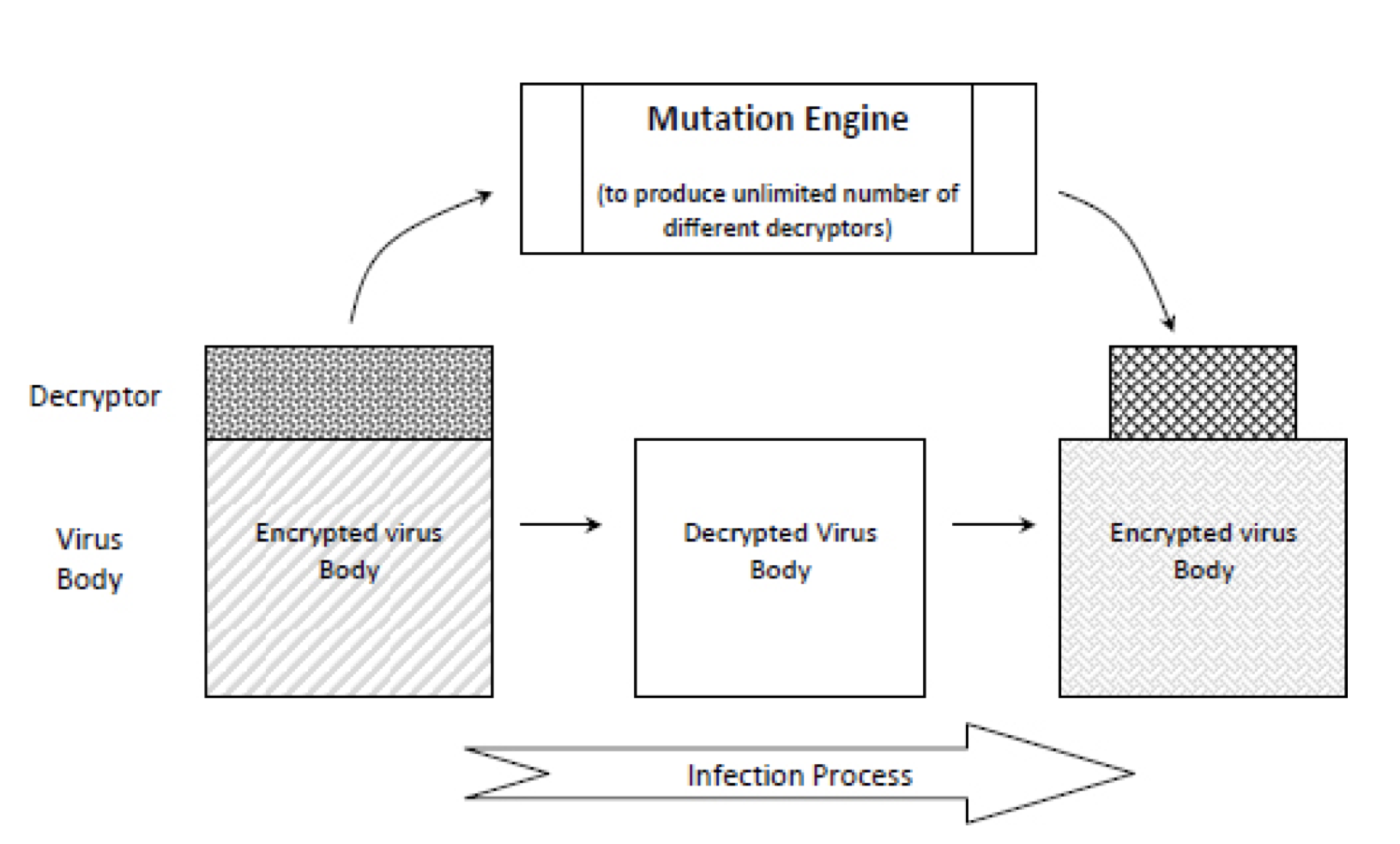}
\caption{Polymorphism virus structure~\cite{rad2012camouflage}.}
\label{f:polymorphism}
\end{figure}

Polymorphism is a more sophisticated method to hide the malicious payload based on malware encryption (Figure \ref{f:polymorphism}). It uses several types of transformations on the decryptor, such as changing the order of instructions with additional jump instructions to maintain the original semantics and permuting the register allocation to deceive anti-virus software~\cite{christodorescu2005semantics}. It also typically injects junk or dead codes to further mutate the decryptor so that it is hard to recognize it~\cite{rad2012camouflage}. However, after enough emulation and a simple string matching algorithm, the underlying encrypted sections of the malware can still be revealed~\cite{rad2012camouflage}. 

\begin{figure}[h]
\centering
\includegraphics[width=\linewidth]{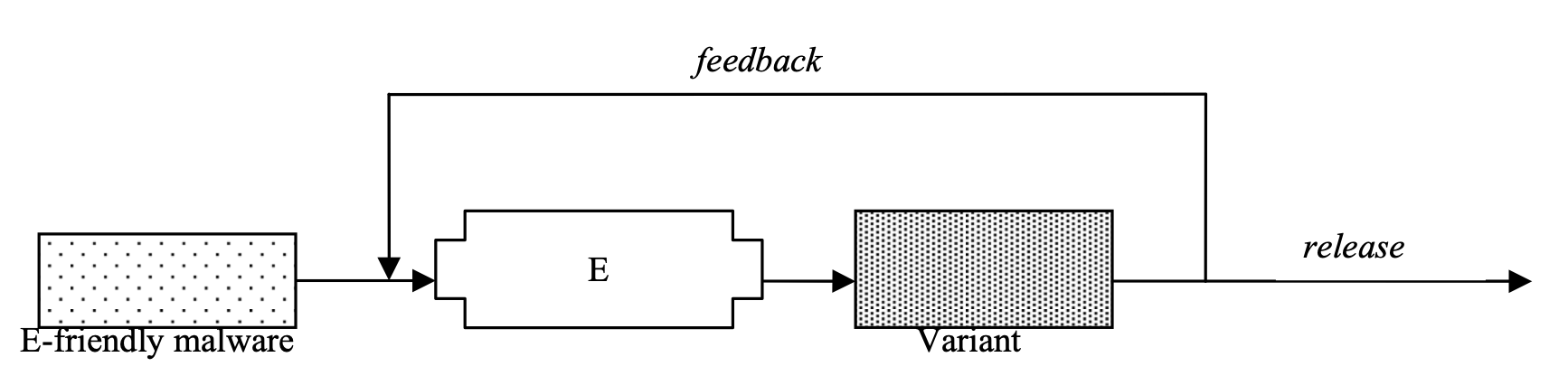}
\caption{Metamorphism structure: a metamorphic engineer is responsible for changing the malware instructions to equivalent ones probabilitically~\cite{rad2012camouflage}.}
\label{f:metamorphism}
\end{figure}

A major drawback of either malware encryption or polymorphism is that it relies on an explicit decryptor to decrypt the malicious payload and execute the malware. This leaves a significant mark on the malware that can be relatively easily detected. On the other hand, metamorphism is a more advanced technique to camouflage the malware without using any encrypted parts. Malware metamorphism is a technique to mutate the malware binary using different obfuscations by a metamorphic engine (Figure \ref{f:metamorphism}). In this way, the attacker changes the syntax of the original malware but keeps the original malware behavior. In particular, metamorphism allows the malware to change its opcode with each execution of the infected program. Alam et al.~\cite{alam2015framework} group some typical obfuscations used in metamorphism into three categories. The first category is the opcode level obfuscation, which includes instruction reordering, dead code insertion, and register renaming~\cite{alam2015framework}. The second category is control flow level obfuscation, which includes changing the order of instructions, and applying branch functions, opaque predicates, jump tables, and exception tables~\cite{alam2015framework}. The last category is obfuscation by self-modifying code, which intends to change instructions during runtime in order to hide malicious payload to avoid reverse engineering and detection by anti-malware software~\cite{alam2015framework}. 

It has been shown that malware metamorphism can easily defeat the signature-based malware detection~\cite{christodorescu2004testing} because signature-based detection is unable to capture the changes of the malware due to dynamic code obfuscation. However, metamorphic malware is usually initiated from known malware, and with the initial knowledge of existing malware, it is still possible to detect malware metamorphism. Zhang et al.~\cite{zhang2007metaaware} proposed a defense to characterize the semantics of the program and perform code pattern matching, based on static analysis of control and data flow of call traces. Alam et al.~\cite{alam2015framework} proposed a metamorphic malware analysis framework that builds the behavioral signatures to detect metamorphic malware in real-time. Chouchane et al.~\cite{chouchane2006using} proposed another method to detect the existence of a metamorphic engineer by checking the likelihood of a piece of code generated by some known metamorphic engines. In addition, the metamorphic malware typically has a significant proportion of binary related to the metamorphic engine, which can be recognized by a more advanced detector, such as a visualization-based malware detector.

% Polymorphism and metamorphism are common techniques under code obfuscation \cite{christodorescu2005semantics}. With metamorphism, the malware can encrypt its malicious payload to hide its intentions and avoid static analysis detection \cite{christodorescu2005semantics,liu2017automatic,islam2013classification}. On the other hand, polymorphism uses several transformations such as changing the order of instructions with additional jump instructions to maintain the original semantics and permuting the register allocation to deceive antiviruses \cite{christodorescu2005semantics}. These techniques will change the software's binary, thus affecting the accuracy of the malware classification process. To the best of our knowledge, there are not many papers in the literature covering the resiliency of malware visualization methods against obfuscation techniques. However, we did find a handful of projects specifically discussing handling code obfuscation methods. 

Go et al. discussed the importance of developing new approaches against polymorphism and metamorphism specifically \cite{go2020visualization}. Their method converts the binary to a grey-scale image and uses the ResNeXt CNN model to build resiliency against such malware camouflage techniques \cite{go2020visualization}. Their paper does not explicitly discuss their method's effectiveness against obfuscation attacks \cite{go2020visualization}. The researchers used the Malimg dataset but did not mention that their dataset contained examples of obfuscation \cite{go2020visualization}. Islam et al. focused more on obfuscation detection by attempting to integrate static analysis with dynamic \cite{islam2013classification}. The paper acknowledges the ease of bypassing static analysis with obfuscation but proposes integrating dynamic analysis using information vectors derived from FLF, PSI and API calls \cite{islam2013classification}. Since obfuscating the features of the code would result in outlier vectors, their approach can detect such attacks \cite{islam2013classification}.

\subsection{Adversarial Machine Learning}
In this section, we introduce adversarial machine learning. In general, adversarial machine learning aims to fool the machine learning model by carefully generating adversarial examples through evasion attacks or polluting the training phase through poisoning attacks. We focus our discussion on evasion attacks leveraged against image classification and malware detection due to their many close similarities to visualization-based malware detection.

\subsubsection{Attacking Image Classification.}

\begin{figure}[h]
\centering
\includegraphics[width=\linewidth]{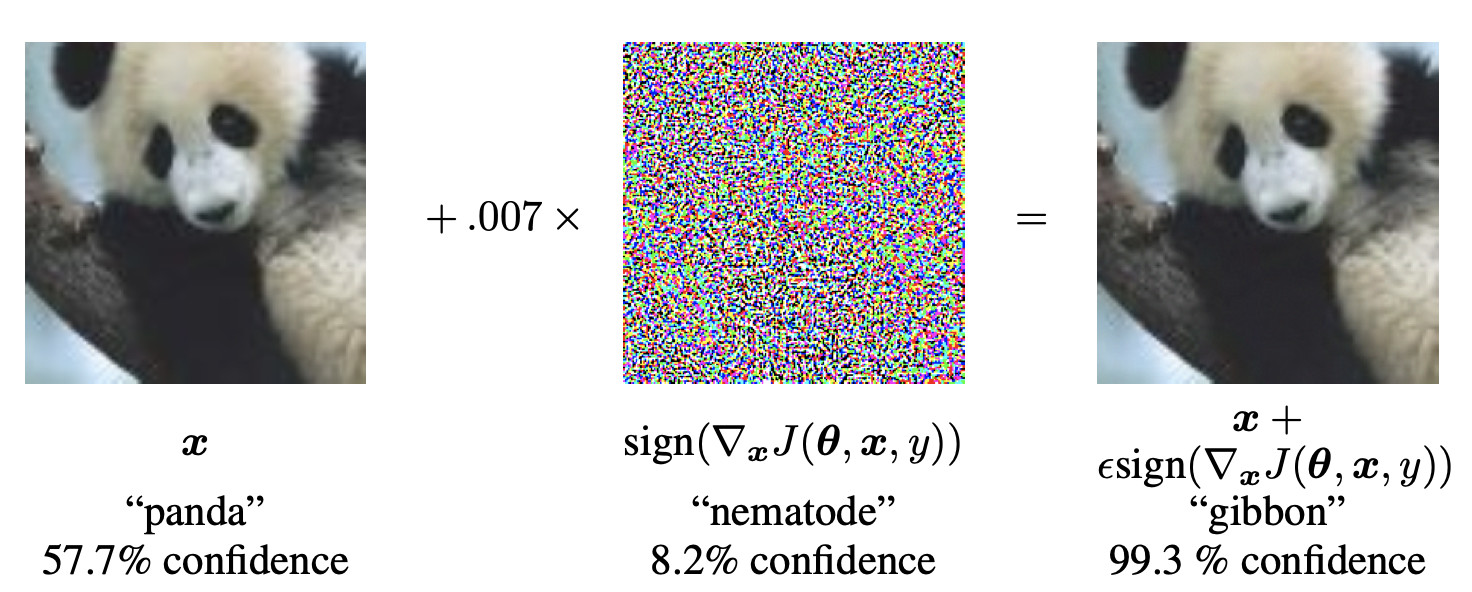}
\caption{Adversarial example generated using FGSM~\cite{FGSM}.}
\label{f:ad_ml}
\end{figure}

Previously, image classification has been used in many applications not related to binary classification. In these fields, multiple attacks have been produced to cause errors in detection. Goodfellow et al.~\cite{FGSM} illustrated the fast gradient sign method (i.e., FGSM) to generate adversarial examples. Figure \ref{f:ad_ml} shows how FGSM is applied to cause the image classifier to mistakenly classify a panda to a gibbon by adding carefully crafted perturbation. In Carlini et al.\cite{carlini2017towards}, it was shown that the addition of random noise to images can significantly reduce the accuracy of classifiers while being imperceptible to the human eye. These attacks can be mitigated to some degree by denoising techniques, as proposed in Liao et al.\cite{Liao_2018_CVPR}. Nevertheless, such mitigation efforts do not fully reverse the perturbations and may introduce more noise accidentally. Furthermore, as shown in Eykholt et al.\cite{Eykholt_2018_CVPR}, classification can also be interrupted by altering small sections of an image, where the image would still be readable by a human eye.

\noindent A notable difference in image classification for binary visualization is that the validator for images lies in code execution instead of human recognition. As a result, perturbations intended for adversarial attacks on binary visualization must continue to function as an unchanged code sample. A code sample that maintains functionality after modification can be said to maintain executability, as the code will execute as intended.

\subsubsection{Attacking Malware Detection.} 

Because malware detection from raw bytes relies heavily on the performance of the machine learning model for classifying the image embedding of the malware (e.g., RGB-colored image), it is also vulnerable to similar attacks against image classification. However, there are more difficulties in generating valid adversarial examples in the malware detection domain than in the image classification domain. The main challenge is to solve the inverse feature-mapping problem~\cite{pierazzi2020intriguing}. Pierazzi et al.~\cite{pierazzi2020intriguing} proposed a novel formalization of problem-space attacks and a novel problem-space attack in the Android malware domain. They used conditional statements that are never executed during runtime to wrap the malicious code payload. In particular, they used opaque predicates to ensure the obfuscated conditions always resolve to false but look legitimate. They showed that their attack against Android is successful against the Drebin classifier~\cite{arp2014drebin} and several SVM-based malware detectors. 

Liu et al. proposed introducing perturbations to the visualized binary to lower the success rate of ML-based malware detectors \cite{liu2019atmpa}. The paper introduced a method that leverages gradient descent and L-norm optimization methods. However, as it changes the image in potentially unexpected ways, it cannot guarantee the executability of the perturbed malware. 

Khormali et al.~\cite{khormali2019copycat} showed simple adversarial examples to bypass visualization-based malware detection. Their attack only attempts rudimentary methods to generate adversarial examples, namely padding and injection~\cite{khormali2019copycat}. It preserves the executability and functionality of the original malware, but it is easy to detect and not scalable.
Kolosnjaji et al. \cite{kolosnjaji2018adversarial} proposed a gradient-based attack to evade ML-based malware detection. Their attack also injects padding bytes to the original malware but does not consider preserving the functionality and does not target visualization-based malware detection. 
Demetrio et al.~\cite{demetrio2020adversarial} proposed a general framework to perform white-box and black-box adversarial attacks on learning-based malware detection by injecting the malicious payload to the DOS header. However, the authors also claimed that their header attacks could be easily patched through filtering process before classification. 
Grosse et al.~\cite{grosse2016adversarial} demonstrated the attack against a deep neural network approach for Android malware detection. They crafted adversarial examples by iteratively adding small gradient-guided perturbation to the malware on application level instead of directly perturbing the binary. They restrict the perturbation to a discrete set of operations that do not interfere with the malware functionality. In addition, they discussed some remedies against their attacks, including feature reduction, distillation, and adversarial training (i.e., re-training the model with the addition of adversarial examples). Their work focused more on attacking  application-level malware detection instead of visualization-based malware detection. 
%Grosse et al.~\cite{grosse2016adversarial} demonstrated the attack against a deep neural network approach for Android malware detection, and discussed a set of remedies against such attacks. They tested several classifiers, which all had an initial accuracy rate of above 90\%, in this field with their attacks, and managed to reached a misclassification, which, in this case, is defined as misleading a classifier to incorrectly guess an input sample, rate of 35\% against every classifier with the attacks the researchers generated \cite{grosse2016adversarial}. However, the work does not seem to focus on preserving the executability of the original malware. 
Sharif et al.~\cite{sharif2019optimization} proposed an optimization-guided attack to mislead deep neural networks for malware detection. Their attack is more invasive in that it changes the reachable code section of the malware. Nevertheless, their attack considers a limited set of transformation types of malware functions and does not target visualization-based malware detection \cite{sharif2019optimization}.
Overall, there is no work proposing robust attacks against the visualization-based malware detection that preserve both executability and functionality of the original malware and are hard to detect through pre-processing. We seek to fill this room in the research space.

\subsection{SoK of Existing Literatures}

Finally, we provide Table \ref{tab:SoK-malicious-package-samples} as our SoK of existing malware visualization techniques and attacks against them. For each reviewed work, we summarize its methodology and list its limitations.

% Attacks against malware detection visualization algorithms, defense against attacks
% Attacks against image classification

% Note to Jingyu: I'm using the same sentences from this paper to describe some of the methodology and limitation. Let me know if that's fine. If you don't understand what I mean, please contact me
\begin{table*}[!h]
\centering
\caption{SoK of reviewed papers: malware visualization and adversarial attacks against malware visualization software.}
\label{tab:SoK-malicious-package-samples}
\begin{tabular}{ |p{2cm}|p{3cm} |p{3.75cm}|p{3.75cm}| }
\hline
Category & Paper & Methodology & Limitations \\
\hline
\multirow{10}{3cm}{\shortstack[l]{Malware\\visualization\\on binary:\\Gray-Scale}} & Han et al.~\cite{han2015malware} & Converts binary to the bitmap image and generates the entropy graph from visualized malware & Hard to classify packed malware binaries\\
\cline{2-4}
& Nataraj et al.~\cite{nataraj2011malware} & Extracts image texture features from visualized malware & Relying on global texture features can be beat by attackers \\
\cline{2-4}
& Xiaofang et al.~\cite{xiaofang2014malware} & Extracts a 64-dimensional feature vector and performs fingerprint matching to identify similar malware & Relying on global image features can be beat by attackers \\
\hline
\multirow{4}{2cm}{\shortstack[l]{Malware\\visualization\\on binary:\\RGB-Colored}} & Fu et al.~\cite{fu2018malware} & Combines entropy, relative section size, and raw bytes to generate an RGB-colored image & Limited to PE format \\
\cline{2-4}
& Han et al.~\cite{han2013malware} & Extracts opcode instruction sequences & Classification is not yet automated \\
\hline
Malware visualization on behavioral features & Shaid et al. \cite{shaid2014malware} & API call monitoring & Does not consider network behavior and does not work directly on malware binary \\
\hline
\multirow{15}{5cm}{\shortstack[l]{Attacking\\malware\\detection}} & Pierazzi et al.~\cite{pierazzi2020intriguing} & General problem-space attack for inverse feature-mapping & Attacks focus on Android malware and are not against neural network-based detector\\
\cline{2-4}
& Liu et al.~\cite{liu2019atmpa} & Gradient descent to perturb binary & Not guarantees executability \\
\cline{2-4}
& Kormali et al.~\cite{khormali2019copycat} & Padding and injection &  Easy to detect and not scalable\\
\cline{2-4}
& Kolosnjaji et al.~\cite{kolosnjaji2018adversarial} & Padding and injection & Does not preserve functionality and does not target visualization-based malware detection \\
\cline{2-4}
& Demetrio et al.~\cite{demetrio2020adversarial} & Injects the malicious payload to DOS header & Easy to patch through filtering process before classification \\
\cline{2-4}
& Grosse et al.~\cite{grosse2016adversarial} & Iteratively adds small gradient-guided perturbations & Only targets Android malware and does not attack on binary level \\ 
\cline{2-4}
& Sharif et al.~\cite{sharif2019optimization} & Manipulates code section guided by an optimization function  &  Considers only a limited set of manipulation types \\
\hline
\end{tabular}
\end{table*}

\section{Robust Adversarial Example Attack against Visualization-Based Malware Detection}
\label{sec:attack}

\begin{figure}[h]
\centering
\includegraphics[scale=0.65]{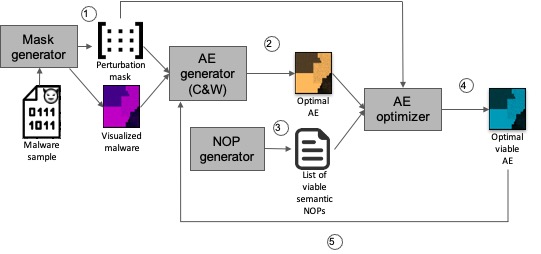}
\caption{The overview of the workflow of the adversarial example generation algorithm.}
\label{f:ae_gen}
\end{figure}

In this section, we focus on the workflow of generating adversarial examples, and we leave the construction of the malware detector to Section \ref{sec:setup}. We have two main goals for our adversarial example attack. Firstly, we aim to find an adversarial example generated from a single malware such that the malware detector will misclassify it as benign software. Secondly, an adversarial example generated in this way must maintain the functionality of the original malware. An overview of the full workflow of our adversarial example generation algorithm is shown in Figure \ref{f:ae_gen}. At a high level, the attack starts with using a mask generator to add empty spaces to instruction boundaries where perturbations are allowed. Then, the adversarial example generator (i.e., \textit{AE generator}) will generate the optimal adversarial example in the image space. To ensure that the original malware functionality is not changed, we use a NOP generator to produce a semantic NOP list and update the optimal adversarial example to the closest matching viable one that preserves malware functionality. If this processed adversarial example is still misclassified as benign, then our attack succeeded. Otherwise, we relaunch the AE generator, starting from the failed adversarial example, creating a new optimal AE, and starting a new iteration. We iterate until we produce a successful adversarial example or we reach a pre-set threshold of iterations. In the following sub-sections, we discuss the details of each component in the workflow.

\subsection{Mask Generator}
The first step of our attack workflow aims at controlling and locating where the perturbations can be added. This step is provided to ensure both executability and robustness to simple pre-processing defenses while maintaining the original semantic operation. The central intuition of this step is to allow additional instructions to be embedded within the code section so that they are not easily distinguishable from the rest of the code section and, in the meantime, ensure that these added instructions do not introduce any changes to the original malware instructions. These additional instructions will serve as empty spaces to allow perturbing the malware in the image representation.

\begin{figure}[!h]
\centering
\includegraphics[width=\linewidth]{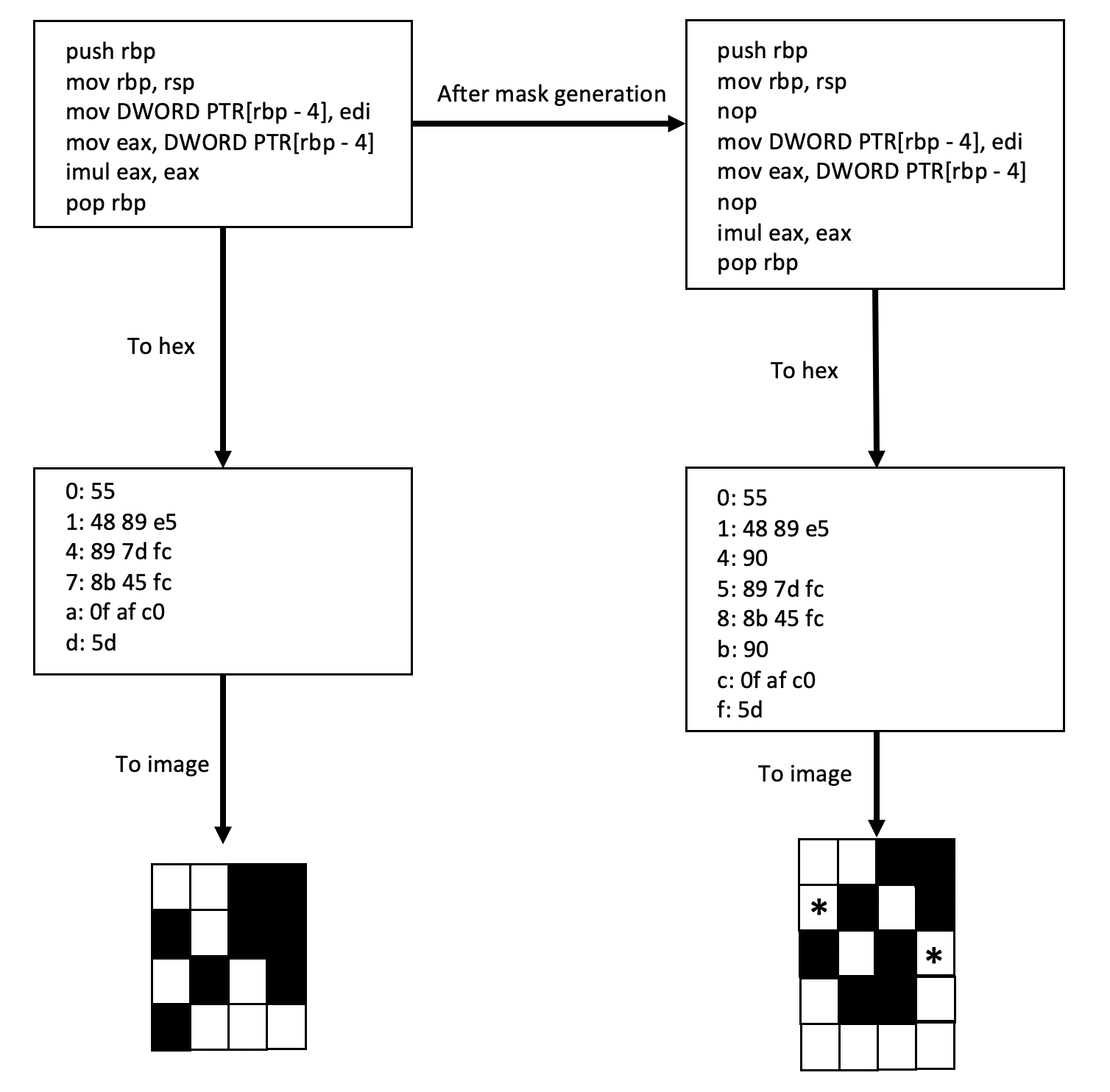}
\caption{Mask generator adds empty space to allow perturbation}
\label{f:mask_gen}
\end{figure}

\noindent To achieve this, we create a \textit{mask generator} (Figure \ref{f:mask_gen}). The algorithm of the \textit{mask generator} is as follows. First, we extract the code section of the malware sample and identify the instruction boundaries. Next, we need to decide the size of each perturbation block, its location, and frequency. The attacker can set these parameters to satisfy certain size limitations of the perturbation relative to the original malware size to make it harder to detect. On the other hand, the attacker can also increase the frequency of the perturbation blocks to make the attack easier. The perturbations can be initialized to random inputs or naive NOPs. After that, these perturbations are inserted into the expected instruction boundaries. In this way, we make sure that the original malware instructions are not changed at all because we never add extra bytes to the middle of the binary of any instruction. With this malware augmented with the perturbation sequences initialized to naive NOPs, we use a binary-to-image converter to represent the malware in the image space. The binary-to-image converter treats the consecutive 8 bits of the binary as a single pixel to build a PNG file. Figure \ref{f:mask_gen} illustrates how our mask generator adds empty space to allow perturbation. In this example, we add a single NOP every two instructions. The image representation of the binary is expanded by two pixels (i.e., the pixel marked with `*') due to the two added nops.

Besides the malware image, the \textit{mask generator} produces a mask in the form of an array of the same dimension as the augmented malware image. The mask flags the locations where perturbations are allowed with ones, while the rest of the array is filled with zeros. We name this mask the \textit{perturbation mask}.

\subsection{AE Generator}
Once the \textit{perturbation mask} and the augmented malware image are generated, we launch a modified version of the CW attack~\cite{carlini2017towards} to generate the optimal adversarial example (i.e., \textit{optimal AE}) in the image space, which is misclassified as benign by the malware detector. The only difference is the application of the \textit{perturbation mask} to further restrict the positions of perturbations. The objective function is given as 
\begin{equation}
    \min ||M\delta||_2 + C\cdot f(x+M\delta)\quad s.t.\quad x+M\delta\in[-1,1].
\end{equation}
Here $M$ is the \textit{perturbation mask}, and $x$ is the augmented malware image produced by the \textit{mask generator}. 
The \textit{optimal AE} is unlikely to correspond to the perturbed malware that maintains the original malware functionality because the CW attack only intends to attack the malware detector, which is modeled as an image classifier. Our objective function does not place additional restrictions to ensure that the perturbations will be either executable or semantic NOPs after being reversed back to the binary, which is required to preserve the malware functionality. One possible approach is to apply the similar idea of ensuring printability for adversarial examples in a real-world attack against image classification. However, the number of semantic NOPs is much larger than the number of printable colors, which will significantly increase the computation overhead of the objective function. On the other hand, the \textit{optimal AE} is a good starting point to guide us to generate a viable adversarial example that keeps the malware functionality. We also avoid complicating the objective function to reduce the runtime overhead to generate the optimal adversarial examples. 

\subsection{NOP Generator}
To generate the viable adversarial example from the \textit{optimal AE}, we need to replace the perturbations introduced by the CW attack with binaries that keep the malware functionality (i.e., semantic NOPs). The replacement algorithm will be discussed in Section \ref{sec:ae_optimizer}. This section shows how we create the semantic NOPs, which are semantically equivalent to a naive NOP, which does not change the values stored in the registers and the function state, such as the stack. We do not use the same strategy in~\cite{pierazzi2020intriguing} that used unreachable conditional statements because that requires careful crafting of obfuscated conditions to avoid being removed by the compiler optimization or the filtering process, which makes the perturbation generation process more complicated. 

\begin{figure}[!h]
\centering
\includegraphics[scale=0.55]{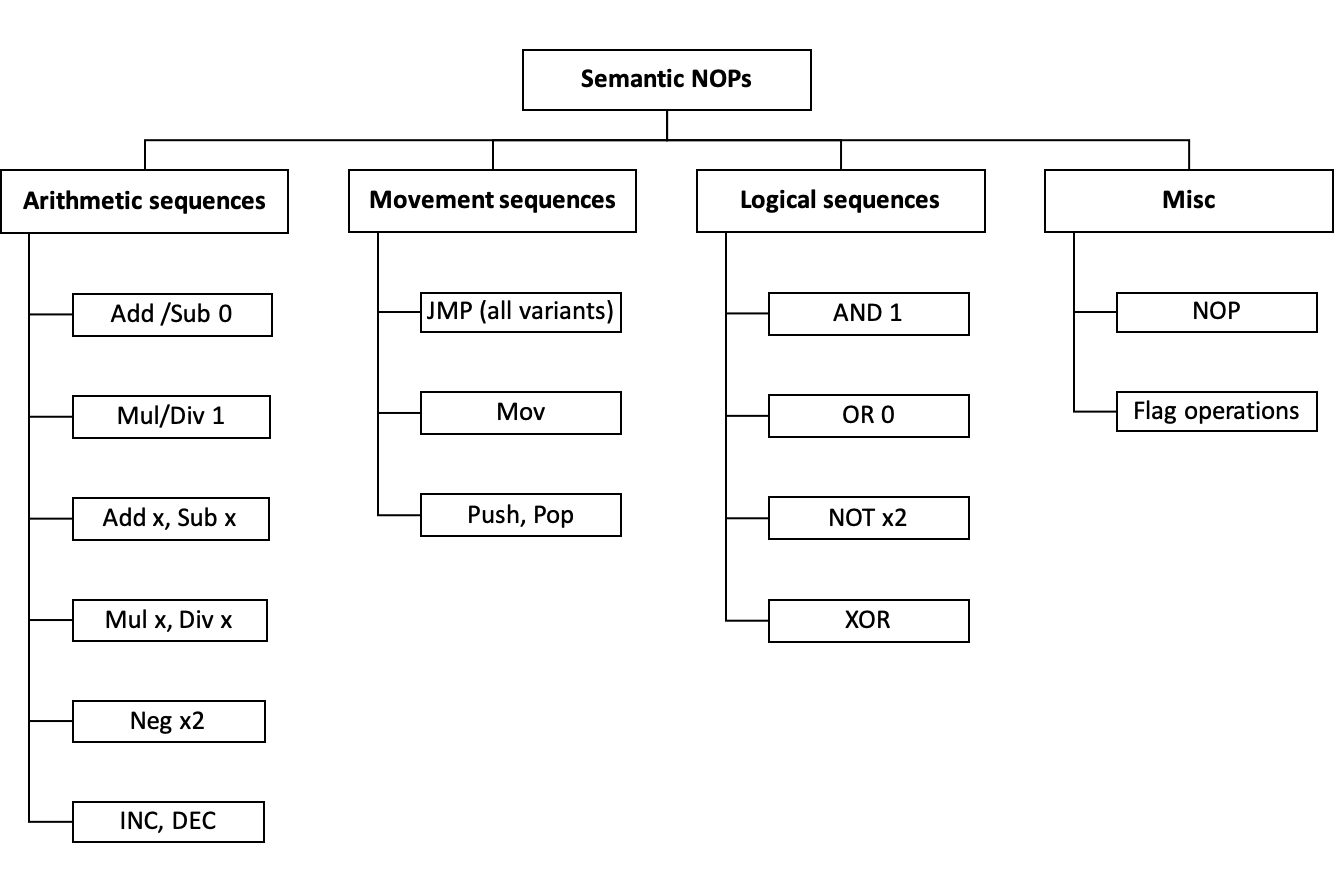}
\caption{Semantic NOP seeds used to construct longer semantic NOPs.}
\label{f:nop}
\end{figure}

\noindent We start by creating short semantic NOPs. We call these short semantic NOPs \textit{semantic NOP seeds}. \textit{Semantic NOP seeds} fall into four categories (Figure \ref{f:nop}):
\begin{enumerate}
    \item Arithmetic sequences: Some neutral sequence of arithmetic operators which can be performed on any register. An example is adding zero to the register.
    \item Movement sequences: Instructions to move the register value back and forth or jump to the defined locations to skip the regions that are not expected to execute. Examples are moving the register value to itself or jumping to the immediate next instruction.
    \item Logical sequences: Some neutral sequence of logical operators which can be performed on any register. An example is ANDing 1 to the register. 
    \item Other miscellaneous sequences: A sequence of simple NOPs or a sequence to change and recover the flags. 
\end{enumerate}

Because the perturbation space is not pre-determined, we do not generate semantic NOPs for any arbitrary size. Instead, we build the \textit{NOP generator} to combine \textit{semantic NOP seeds} to make longer semantic NOPs that can fit larger perturbation spaces. For instance, if the \textit{NOP generator} is asked to compute 3-byte semantic NOPs and the naive NOP (i.e., byte 90 in heximal) is one of the \textit{semantic NOP seeds}, then it can combine three NOPs. Given the expected size of each perturbation block, the \textit{NOP generator} produces a list of semantic NOPs of the given size.

It is necessary to keep the byte length of \textit{semantic NOP seeds} as small as possible so as to improve the ability for the \textit{AE optimizer} to produce a viable adversarial example that maintains the malware functionality. However, too much restriction on the byte length also limits our ability to generate enough \textit{semantic NOP seeds}. In our design, we pick the minimum size of a \textit{semantic NOP seed} as one byte (i.e., the byte length for a naive nop) and the maximum size of a \textit{semantic NOP seed} as eight bytes. We provide our example \textit{semantic NOP seeds} and the corresponding byte size in Table \ref{tab:nop_size}. We admit that this is a simple design choice and is far from the optimal selection. We also do not comprehensively list all the possible operations given the operation type, which can make the set of \textit{semantic NOP seed} too large. However, our evaluation results reveal that this selection is already enough for us to generate adversarial examples for malware that maintains functionality with high success rate. 

\begin{table}[!h]
\caption{Example of semantic NOP seeds and their byte length.}
\label{tab:nop_size}
\centering
\begin{tabular}{|l|l|l|}
\hline
Operation Type & Example in Hex & Byte Length \\
\hline
Nop & 90 & 1 \\
Move register to itself & 89c0 & 2 \\
Jump to next instruction & 7700 & 2 \\
Push and pop & 5058 & 2 \\
Not and not & f7d0f7d0 & 4 \\
Add/subtract 0 & 9c83c0009d & 5 \\
Logical AND with 1 & 9c83e0ff9d & 5 \\
Logical OR with 0 & 9c83c8009d & 5 \\
Logical XOR with 0 & 9c83f0009d & 5 \\
Negate and negate & 9cf7d8f7d89d & 6 \\
Increment and decrement & 9cffc0ffc89d & 6 \\
Add x and subtract x & 9c83c00183e8019d & 8 \\
\hline
\end{tabular}
\end{table}

\subsection{AE Optimizer}
\label{sec:ae_optimizer}
In this step, we build a module, the \textit{AE optimizer}, which produces a viable adversarial example that maintains the original malware functionality. The \textit{AE optimizer} takes in the \textit{perturbation mask}, the \textit{optimal AE} generated from the CW attack, and the list of semantic NOPs produced by the \textit{NOP generator}. Next, the \textit{AE optimizer} locates the allowed positions for perturbations in the \textit{optimal AE} using the \textit{perturbation mask}. Subsequently, the Euclidean distance between the instruction sequences in the allowed perturbation spaces and the semantic NOPs is calculated using the following equation: 
\begin{equation}
d\left( p,q\right)   = \sqrt {\sum _{i=1}^{n}  \left( q_{i}-p_{i}\right)^2 }
\label{eq:1}
\end{equation}
Here $p$ and $q$ are the generated instruction sequence and the semantic NOPs. This process identifies the semantic NOPs closest to each of the sequences in the allowed perturbation space of the \textit{optimal AE}. In the current implementation, this process is done sequentially for each perturbation block; however, it can be easily parallelized to improve the runtime performance. After that, the semantic NOPs with the minimum distance are used to replace the perturbation blocks in the \textit{optimal AE}. The new adversarial example is called the \textit{optimal viable AE.}

Finally, we pass the \textit{optimal viable AE} to our malware detector for classification. If it is classified as benign, we stop the process because we have already produced a successful adversarial example. If it is correctly classified as malware, it will be used as the starting point for another iteration of the CW attack, and the process is repeated. We expect that starting from a failed optimal viable AE can direct us better to the successful optimal viable AE. However, it is possible that the AE can be stuck in the local optimum. Another more general approach is to start over the whole process again from the visualized malware augmented with randomly initialized semantic NOPs.

\section{Evaluation}
\label{sec:eval}

\subsection{Experiment Setup}
\label{sec:setup}

\subsubsection{Malware Detection Model.}

\begin{figure}[!h]
\centering
\includegraphics[width=1.1\linewidth]{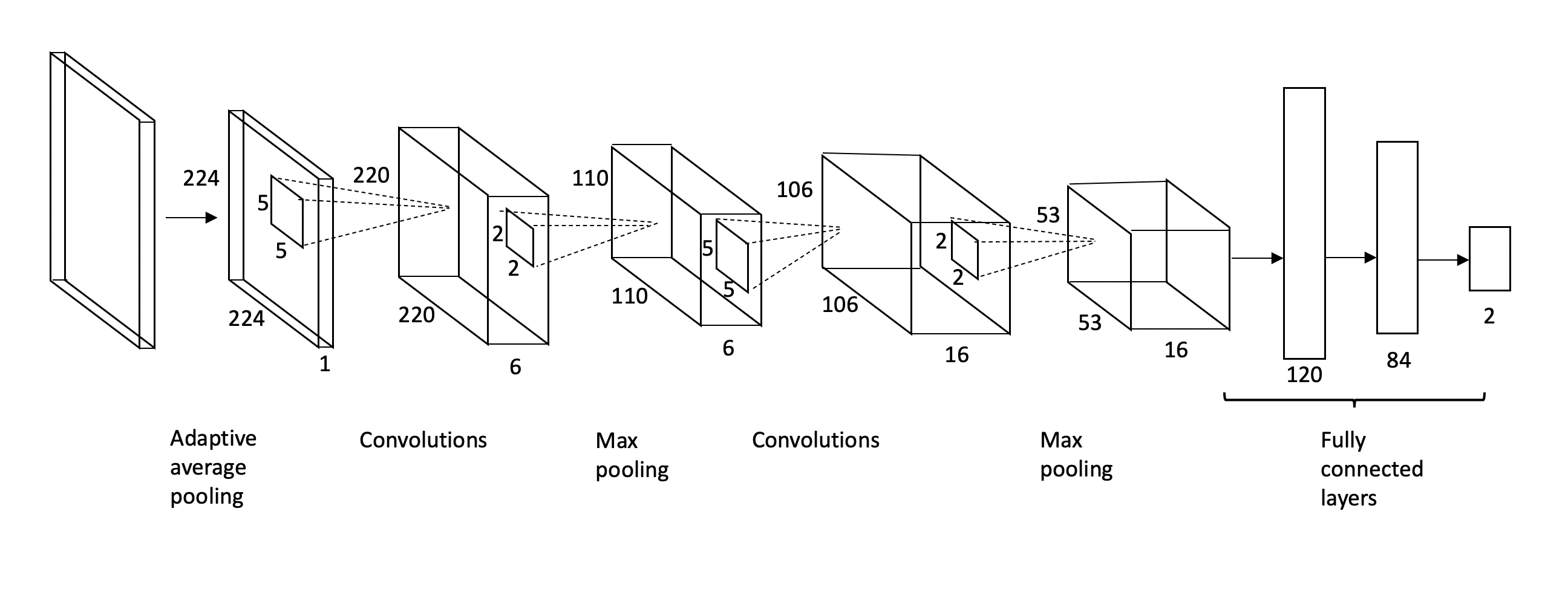}
\caption{Malware detection neural network model structure.}
\label{f:nn}
\end{figure}

We use a CNN-based malware detector as our attack target. Because there is no open-source code for the visualization-based malware detector, we build our own CNN by following a similar structure from previous work. Specifically, we re-built the structure described in~\cite{khormali2019copycat,kolosnjaji2018adversarial}. Our CNN is composed of an adaptive average pooling layer to handle inputs of different dimensions, two consecutive convolutional layers with max-pooling, and three fully connected layers. We use ReLU as the activation function (Figure \ref{f:nn}).

\subsubsection{Dataset.}

We build our dataset by combining one public malware dataset and one public benign software dataset. The malware dataset is from UCSB's public malimg dataset~\cite{malimg}, which has a collection of around 9500 malware formatted as PNGs. The benign dataset is a collection of around 150 benign software from the Architecture Object Code Dataset~\cite{AOCDdataset}. The software was chosen from the AOCD with various functionalities to ensure that the classifier did not learn latent features representative of the type of benign code.

\subsubsection{Model Training.}

A relatively small subset of the malimg dataset is used to acquire a $50\%$ split between malware and benign images in the training data. This subset was retrieved randomly from the malimg dataset, and examples were chosen without considering their corresponding malware family. In doing this, we prevent the classifier from learning any latent features from any particular malware family and improve its accuracy for any malware class in the validation set. Validating this model's accuracy, we are able to confirm the classification accuracy of state-of-the-art models, with a $100\%$ accuracy on identifying benign software and a $99\%$ accuracy on correctly identifying malware samples.

\subsubsection{Attacking the Malware Detector.}

We evaluate our attack using the malware from the malimg dataset~\cite{malimg}. For each malware, we augment it with empty spaces per instruction initialized with eight naive NOPs. If the \textit{AE generator} fails to produce the \textit{optimal AE}, we consider it an attack failure. If the \textit{AE generator} can produce the \textit{optimal AE}, we run the \textit{AE optimizer} to replace perturbations with the closest semantic NOPs. We set the iteration threshold to be ten. If the \textit{optimal viable AE} is generated within the ten iterations of the CW attack and AE optimization and can be successfully misclassified as benign, we view it as a successful attack.

\subsection{Results}

% TODO: Reviewer 1: reason for failure of 2 malware
% Idea: Discuss oscillation between objective function. To improve, try random initialization in the first step.

We first evaluate our attack on the malware family, ``Dialplatform.B'', which contains 174 malware samples. Malware in this family has moderately large code section, which gives us more flexibility to generate adversarial examples in a short time. The corresponding image of each malware is either $216\times 64$ or $432\times 64$ in pixels. The average number of instructions in the code section for each malware is 622. All the malware is classified as malware by our malware detector. 

We successfully generate adversarial examples for 172 malware out of the total 174 malware (98.9\%).
For the two malware that we fail to generate an adversarial example for, our algorithm workflow oscillates around a local optimum, so our iteration of CW attack and AE optimization (steps 2, 4, 5 in Figure \ref{f:overview}). There are two potential methods to avoid the local minimum oscillation issue. First, we can initialize the empty spaces between instructions with random semantic NOPs. Randomized initialization has already been applied frequently to solve similar local optimum oscillation problems before. Second, in the AE optimizer step (Section \ref{sec:ae_optimizer}), we could pick a sub-optimal viable adversarial example if we detect local optimum oscillation. In this way, our algorithm can break the local optimum while also searching for a valid adversarial example.
% it is more likely that our algorithm can break the local optimum oscillation range and is still in good progress of finding a valid adversarial example. 

%TODO: Reviewer 2: space expansion tradeoff

\begin{figure}[!h]
\centering
\includegraphics[width=\linewidth]{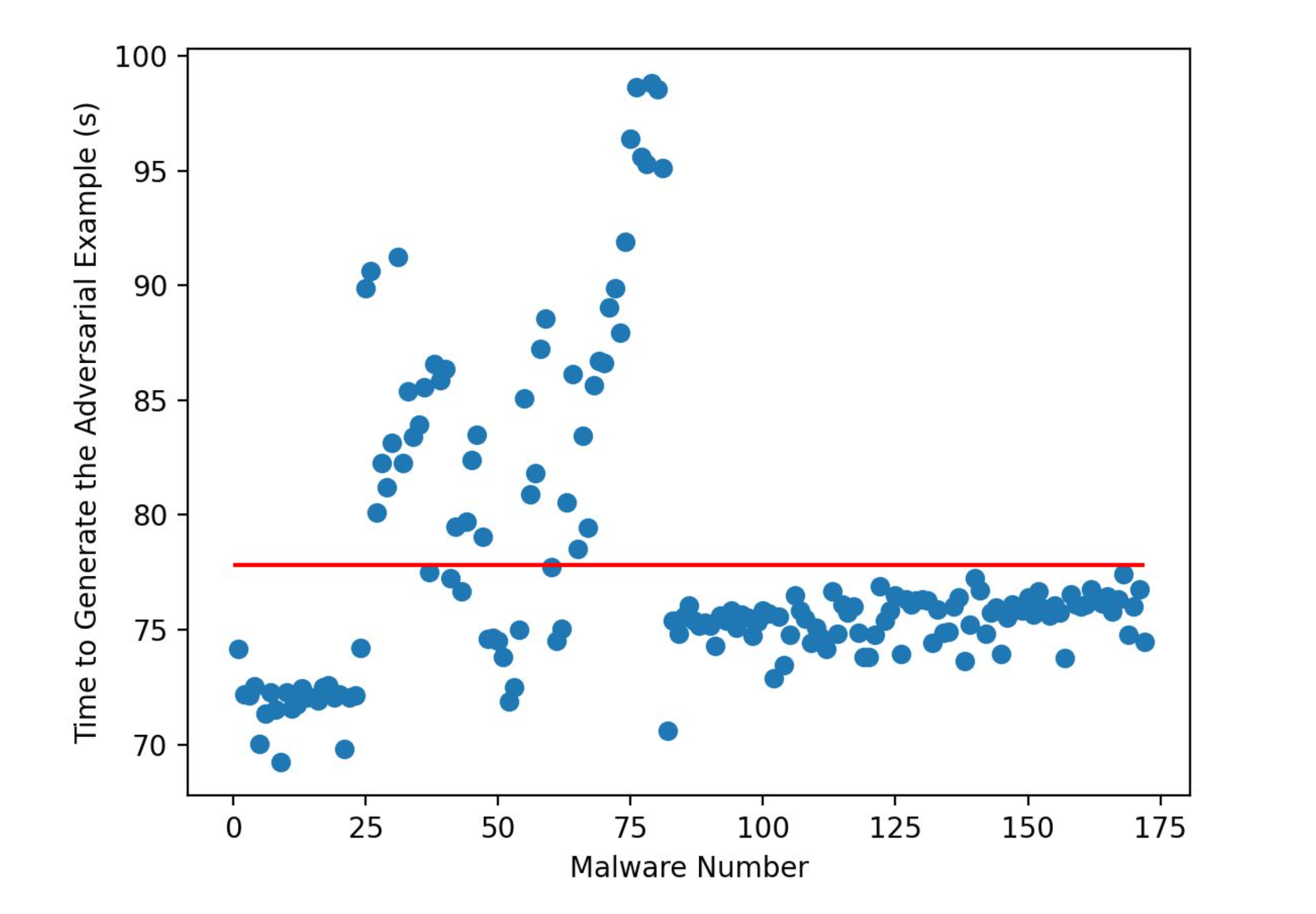}
\caption{The runtime evaluation for the adversarial example attack.}
\label{f:time}
\end{figure}

\begin{figure}[!h]
\centering
\includegraphics[width=\linewidth]{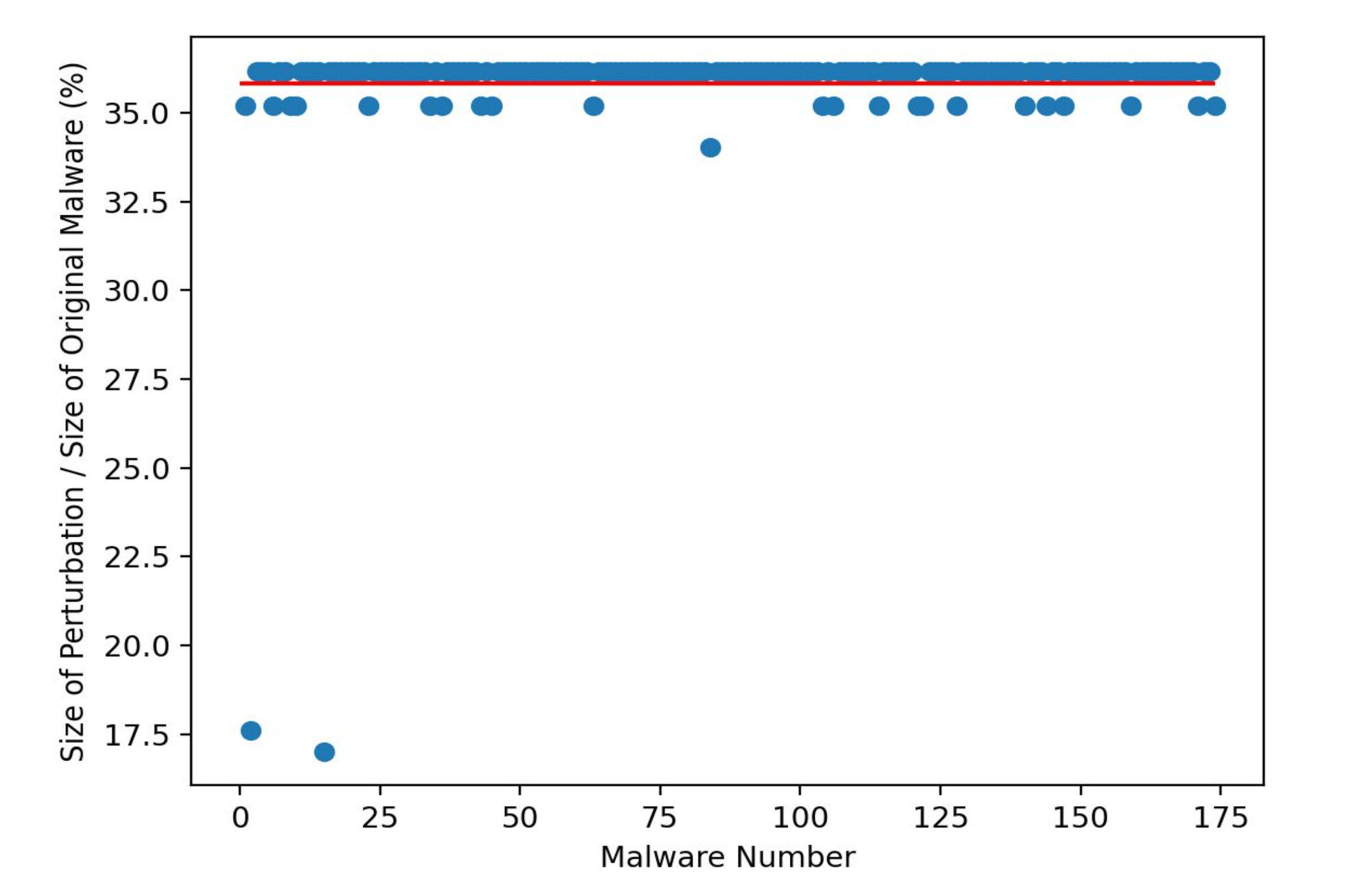}
\caption{The size of the added perturbation of the adversarial example attack.}
\label{f:size}
\end{figure}

For all of the adversarial examples, the functionality of the original malware is preserved by construction since we only instrument the original binary with semantic NOPs at instruction boundaries. All of the adversarial examples can be generated within five iterations of the CW attack and AE optimization. The running time to generate the adversarial example is given in Figure \ref{f:time}. On average, it takes 77.8 seconds to generate the adversarial example for the malware with around 600 instructions, and the time overhead is in AE optimization step. The expansion rate of the original malware due to augmentation is shown in Figure \ref{f:size}. On average, the perturbation size is 35.82\% of the size of the original malware as seen from Figure~\ref{f:size}. We argue that though the expansion rate is high, the added perturbation is still hard to filter out due to the flexibility of our semantic NOPs. The attackers can even build a much larger semantic NOP set than our current version. They can also consider using dead code branching. In this way, unless the users know the semantic NOPs and the initial size of the malware, it is hard to filter out our attacks. 

We achieve similar results for other malware families with similar sizes of the code section. On the other hand, our algorithm does not work well for malware with very small code sections. We argue that the spaces to add semantic NOPs are too few and too small to allow enough perturbation to cause misclassification due to the small code section. We expect that enlarging the mask size can improve the attack success rate, but this might defeat the attacker's purpose for distributing small and easy-to-transmit malware. Another potential way is to combine perturbations in code sections with those in other sections, such as data sections. However, we argue that keeping executability can be hard and naive padding is relatively easy to filter out. We leave attacks for malware with a small code section as future work. 

% TODO: Reviewer 1: reason of 18.2% failure.
% Idea: Reason should also be oscillation. We can also mention random initialization improvement. We need to explicitly admit our shortcomings: though it is possible, a naive approach without parallelism will introduce too long time. Future work include implementing parallelism and run it on a larger dataset. This experimental mainly shows a proof-of-concept naive attack and true effectiveness needs further work to analyze.

\noindent To further test the end-to-end methodology for generating adversarial examples for malware with a larger number of instructions, we also run the same experiment with 11 malware from the ``Lolyda.AA3'' class. Each malware contains at least 200,000 instructions. We achieve an attack success rate of 81.8\% (i.e., 9 out of 11). The two failures are also due to local optimum oscillation issues that we face when generating adversarial examples for malware in ``Dialplatform.B'' family. We expect random initialization and sub-optimal AE optimizer can solve the problem and leave it to future work. On the other hand, the major problem for generating adversarial examples for large malware is the time overhead. In our experiments, the time overhead reaches about six hours to generate a single adversarial example. As a proof-of-concept attack and assuming the attacker can tolerate this time overhead, our attack can still work properly, but the practicality of the attack in real life is still under question considering the size overhead added to the original malware as well. We expect parallelism can improve the running time. Specifically, our current AE optimizer finds viable semantic NOP sequences for each empty space defined by the mask sequentially. However, each empty space is independent of the other if we perturb it with self-contained NOP sequences. Therefore, parallelization can be easily applied in the AE optimization step. We leave its implementation and evaluation as future work.

Our high attack success rate is a conservative representation of how vulnerable visualization-based malware detection can be to the adversarial examples. In our experiment, we set each perturbation block to be small (i.e., 8 bytes). A more powerful attacker who would like to risk being detected more can further increase the size of each perturbation block so that the attack can be easier to launch.

\section{Discussion}
\label{sec:discussion}
\subsection{Limitations}
While the proposed attack can successfully find adversarial examples with a high success rate, due to the nature of the optimization algorithm, the time necessary to find the \textit{optimal viable AE} increases drastically with the size of the original malware. In our evaluation, we performed adversarial example attacks on the ``DialPlatform.B'' malware family, where each malware image is of dimension $216\times 64$ or $432\times 64$ with an average of 622 instructions. Since the images are of reasonably small dimensions, the potential perturbation spaces are fairly limited. As a result, an adversarial example can be generated in less than two minutes. However, for larger malware, as in the ``Lolyda.AA3'' family, producing a single optimal adversarial example can take a few hours. As each perturbation block can be independently replaced with the closest semantic NOPs, we expect parallelization to improve the running time.

In addition, our attack only adds instruction to the malware's code sections. Therefore, when the code section for the original malware is small, it will be hard to provide enough perturbations to cause the classifier to misclassify the malware image. Similar issues occur when the data section is much larger than the code section. One possible approach to find an easier way to generate perturbation is to find out the hot zone for the machine learning detector on the malware in the first place and then only adding enough semantic NOPs to these locations. Another approach to solving this challenge is to design a mechanism to perturb the data section without affecting the original functionality of the malware.

There are some potential defenses against general attacks to malware detection. Tong et al.~\cite{tong2019improving} proposed a method to boost the robustness of feature-space models by identifying the conserved features and constraining them in adversarial training. Their defense is evaluated only on PDF malware, and we plan to evaluate further the defense on more malware families and more diverse model types. 

\subsection{Future Work}
The proposed attack algorithm is a proof-of-concept and a work-in-progress, and there are several research directions we would like to explore as future work: 
\begin{enumerate}
    \item In the current version of the attack, the size and frequency of the added perturbations are chosen beforehand by the attacker. In the development of the attack, we would like to explore the possibility of adding these two parameters (i.e. size and frequency of the perturbations) to the optimization process. We speculate that this can improve the performance of our AE generation algorithm. First, it can lead to faster convergence into a viable AE. Additionally, it can also lead to a smaller perturbation size that is customized for each malware sample. 
    \item Another avenue of improvement to our baseline algorithm is in speed performance as the baseline algorithm does not leverage any speed optimizations. Our AE generation algorithm can allow batching and parallelization to enhance the speed performance. First, we plan to batch instructions to be processed at once instead of processing individually. Additionally, since Euclidean distance calculations do not have to be done sequentially we plan to calculate all the distances in parallel.
    \item Another avenue that we would like to explore is the intrinsic properties of NOPs. Specifically, we would like to study the difference between NOPs and whether some NOPs are better than others. Additionally, we would like to draw from the field of software engineering in creating semantically similar code using modifications that do not change the semantics at a higher-level language (e.g. adding empty loops or if statements that would never be true).
    \item In our current implementation, we restrict adding perturbation to the code section. In the algorithm, we would like to explore the effects of adding perturbations to other sections and understanding its effects on executability, robustness to pre-processing, and maintaining semantic operations.
    \item We plan to perform a comprehensive comparison between our proposed attack with the state-of-the-art attacks with respect to speed, stealthiness, and deviation from the original binary. Additionally, we want to evaluate our attack's success against the defense and detection mechanisms available (e.g. VirusTotal).
    \item To have an end-to-end solution we would like to add a module that checks executability of each of the binary (e.g., using \emph{execve} command).
    \item Our current work has already revealed that malware detection based on binary visualization can be beaten by an adversarial example with a high success rate. In the meantime, the attacker can also maintain the functionality of the original malware. Therefore, our work motivates future directions to propose defenses against our attack. Previous defenses against adversarial examples, such as defensive distillation~\cite{papernot2016distillation} and adversarial training~\cite{goodfellow2014explaining}, does not usually focus on real-world tasks, such as defending malware detection models. Whether these defenses are effective in this use case is worth exploring. 
\end{enumerate}

\section{Conclusion}
In this work, we provide a literature review on existing techniques and attacks for malware detection. We summarize their limitations, and propose a novel end-to-end method for generating adversarial examples against visualization-based malware detection. We design our attack workflow to ensure that the malware sample remains executable while maintaining its original functionality. Additionally, we ensure that the added perturbation is robust against pre-processing by inserting semantic NOPs in the reachable code section. We construct a dataset that includes both malware and benign software samples and use it to build an visualization-based ML malware detector that achieves high accuracy. Next, we design a workflow that generates semantic NOP sequences and use them to construct viable adversarial examples. Our results show that it is possible to successfully generate adversarial examples that can bypass a highly accurate visualization-based ML malware detector while maintaining executability and without changing the code operation. Our work motivates the design for more robust visualization-based malware detection against carefully crafted adversarial examples.

\bibliographystyle{splncs04}
\bibliography{mybibliography}
\end{document}